\documentclass[twocolumn,prl,showpacs,amsmath,amssymb,superscriptaddress,floatfix]{revtex4-1}

\usepackage{bm}
\usepackage{graphicx}

\newcommand{\be}{\begin{equation}}
\newcommand{\ee}{\end{equation}}
\newcommand{\bea}{\begin{eqnarray}}
\newcommand{\eea}{\end{eqnarray}}
\newcommand{\ba}{\begin{array}}
\newcommand{\ea}{\end{array}}

\begin{document}

\title{Symmetries of multifractal spectra and field theories of Anderson localization}

\author{I.~A.~Gruzberg}
\affiliation{The James Franck Institute, The University of Chicago, Chicago, Illinois 60637, USA
}

\author{A.~W.~W.~Ludwig}
\affiliation{Department of Physics, University of California, Santa
Barbara, California 93106, USA
}

\author{A.~D.~Mirlin}
\affiliation{Institut  f\"ur Nanotechnologie,  Karlsruhe Institute of
  Technology, 76021 Karlsruhe, Germany
}
\affiliation{Institut f\"ur Theorie der kondensierten Materie,
Karlsruhe Institute of Technology, 76128 Karlsruhe, Germany
}
\affiliation{Petersburg Nuclear Physics Institute,
 188300 St.~Petersburg, Russia
}

\author{M.~R.~Zirnbauer}
\affiliation{Institut f\"ur Theoretische Physik, Universit\"at zu
  K\"oln, Z\"ulpicher Strasse 77, 50937 K\"oln, Germany}

\begin{abstract}

We uncover the field-theoretical origin of symmetry relations for multifractal spectra at Anderson transitions and at critical points of other disordered systems. We show that such relations follow from the conformal invariance of the critical theory, which implies their general character. We also demonstrate that for the Anderson localization problem the entire probability distribution for the local density of states possesses a symmetry arising from the invariance of correlation functions of the underlying nonlinear $\sigma$ model with respect to the Weyl group of the target space of the model.

\end{abstract}


\maketitle

More than half a century after its discovery, Anderson localization \cite{AL50} remains a vibrant research field. One of the central research directions is the physics of Anderson transitions \cite{evers08}, including metal-insulator transitions and transitions of quantum Hall type (i.e.\ between different phases of topological insulators). Apart from electronic conductors in semiconductor structures, experimental realizations include localization of light \cite{wiersma97}, cold atoms \cite{BEC-localization}, ultrasound \cite{faez09}, and optically driven atomic systems \cite{lemarie10}. On the theory side, the field received a strong boost through the discovery of unconventional symmetry classes and the development of a complete symmetry classification of disordered systems \cite{altland97, zirnbauer96, evers08, heinzner05}. These classes are characterized by additional particle-hole and/or chiral symmetries. Examples include disordered superconductors and graphene.

A remarkable property of Anderson transitions is the multifractality of wave functions, describing their strong fluctuations at criticality. Specifically, in $d$ dimensions, the wave function moments show anomalous multifractal (MF) scaling with respect to the system size $L$,
\be
\label{e1}
L^d \langle |\psi({\bf r})|^{2q} \rangle \propto L^{-\tau_q}, \qquad \tau_q = d(q-1) +  \Delta_q,
\ee
where $\langle \ldots \rangle$ denotes the disorder average and $\Delta_q$ are anomalous MF exponents distinguishing the critical point from a simple metallic phase (where $\Delta_q \equiv 0$). Closely related is the scaling of moments of the local density of states (LDOS) $\rho(r)$,
\be
\label{e2}
\langle \rho^q \rangle \propto L^{-x_q} , \qquad x_q = \Delta_q + qx_\rho,
\ee
where $x_\rho \equiv x_1$ controls the scaling of the average LDOS, $\langle\rho\rangle \propto  L^{-x_\rho}$. First steps towards experimental determination of MF spectra have been made recently \cite{faez09, lemarie10, richardella10}.

In Ref.~\cite{mirlin06}, an exact symmetry for MF exponents
\be
\label{e3}
\Delta_q = \Delta_{1-q},
\ee
was derived for any critical system in the conventional Wigner-Dyson (WD) classes as a consequence of a more general relation \cite{mirlin94, fyodorov04} for the LDOS distribution function (and thus, for the LDOS moments),
\be
\label{e4}
{\cal P}(\rho) = \rho^{-3}{\cal P}(\rho^{-1}), \qquad \langle \rho^q \rangle = \langle \rho^{1-q} \rangle.
\ee
Equation (\ref{e4}) is exact at the level of the nonlinear $\sigma$ model and is fully general otherwise; i.e., it is equally applicable to metallic, localized, and critical systems. While in general $\sigma$ models are approximations to particular microscopic systems, Eq. (\ref{e3}) is exact in view of universality of the critical behavior \cite{mirlin06}. See also \cite{monthus09}.

The goal of the present work is to reveal the field-theoretic basis underlying the symmetry relations (\ref{e3}) and (\ref{e4}), and to generalize them to a broader class of systems. First, using arguments arising from conformal invariance at the transition, we show that relations analogous to (\ref{e3}) are valid for a wide class of critical points in disordered systems (that need not be Anderson transitions) characterized by multifractality. Second, focusing on the problem of Anderson localization, we demonstrate that Eqs.~(\ref{e3}) and (\ref{e4}) are manifestations of the Weyl group symmetry of the nonlinear $\sigma$-model theory. Finally, we use this to generalize  Eqs.~(\ref{e3}) and (\ref{e4}) to the unconventional symmetry classes C and CI (in the notation of Refs. \cite{altland97, zirnbauer96}). At the end we describe applications of our results to a number of specific disordered systems.

We begin by presenting a general argument based on (global) conformal invariance of a critical system in $d$ dimensions. Consider a system at criticality characterized by operators ${\cal O}_q$ representing moments of an observable of interest. In the case of Anderson localization on which we focus, this observable is the LDOS, and ${\cal O}_q$ corresponds to $\rho^q$, but one can apply the argument to a broader class of systems. Generically, the spectrum of the scaling dimensions $x_q$ of the operators ${\cal O}_q$ is convex, $x_q'' < 0$ (primes denote derivatives with respect to $q$), satisfies $x_0 = 0$, and becomes negative at sufficiently large (positive or negative) values of $q$ \cite{evers08}. Therefore, there is a single point $q_*$ (in addition to $q=0$) such that $x_{q_*}=0$. Let us show that $q_* >0$. Indeed, it is easy to see that the derivative $x'_0 \equiv (dx_q/dq)_{q=0} = \alpha_0 - d + x_\rho$, where $\alpha_0 \equiv (d \tau/ d q)_{q=0} = \tau'_0$ controls the scaling of a typical wave function amplitude, $|\psi^2|_{\rm typ}\sim L^{-\alpha_0}$. Normalization of the wave function implies that $\alpha_0 > d$ \cite{evers08}. In the WD symmetry classes where $x_\rho=0$, this guarantees that $x'_0 > 0$. In the unconventional classes, $x_\rho$ may be nonzero, with either sign. However, generalizing the conformal invariance argument from Ref.~\cite{obuse10}, we can show that $\alpha_0 - d + x_\rho$ determines the typical localization length in a quasi-1D geometry, implying again that $x'_0 > 0$. It follows immediately that $q_*>0$.

According to the definition of the operators ${\cal O}_q$, their operator product expansion has the form \cite{duplantier91}
\be
\label{e5}
{\cal O}_{p}(r_1) {\cal O}_{q}(r_2) \sim  |r_1-r_2|^{x_{p+q} - x_p - x_q}{\cal O}_{p+q}\Big({r_1+r_2\over 2}\Big).
\ee
In general, the operator $ {\cal O}_{p+q}$ has nontrivial scaling with the system size, $\langle {\cal O}_{p+q} \rangle \propto L^{-x_{p+q}}$; the existence of negative scaling dimensions distinguishes disordered critical points from conventional ``unitary'' conformal field theories. However, in the case of $p = q_* -q$, the operator on the right-hand side of (\ref{e5}) has zero scaling dimension  $x_{q_*}=0$. Therefore, the correlation function $\langle {\cal O}_{q_*-q}(r_1) {\cal O}_{q}(r_2) \rangle $ does in fact not depend on the system size  $L$ (i.e.\ on the infrared regularization of the theory). This allows us to apply the standard argument from conformal invariance \cite{CFT} according to which the nonvanishing two-point correlation function appearing in the expectation value of (\ref{e5}) implies that the dimensions of the nonderivative \cite{Non-Derivative} operators ${\cal O}_{q_*-q}$ and ${\cal O}_{q}$ are equal; i.e.
\be
\label{e6}
x_q = x_{q_* -q}.
\ee
This is the generalized symmetry relation for MF exponents. Note that in the case of Anderson transitions the symmetry holds in general for the LDOS exponents $x_q$ (the scaling dimensions of local field operators) rather than for the wave function exponents $\Delta_q$. (For the WD classes $x_\rho=0$, so that $x_q=\Delta_q$.) The symmetry point $q_*/2$ remains unspecified by the above argument.

Below we show that a stronger symmetry relation between moments of the LDOS, valid for nonlinear $\sigma$ models describing disordered systems of non-interacting fermions in the WD classes (A, AI, AII) as well as those in the Bogoliubov--de Gennes classes with preserved spin rotation invariance (C, CI),
\be
\label{e7}
\langle \rho^q \rangle = \langle \rho^{q_*-q} \rangle,
\ee
has a group-theoretic origin. As a result, the symmetry point $q_*/2$ for these systems is determined solely by the symmetry class and is independent of further details of the problem (e.g., spatial dimensionality, presence or absence of topological order, and whether the system is in a metallic, insulating, or critical phase). Before presenting the proof, let us first ask the following question: Assuming that we know that (\ref{e7}) holds with $q_*$ depending on the symmetry class only, what is a simple way to find $q_*$? It turns out that it suffices to analyze a zero-dimensional $\sigma$ model, equivalent to a random matrix (RM) model.

To see this, let us consider a RM model and introduce a small broadening $\delta \ll \Delta$ (where $\Delta$ is the mean level spacing) for all levels. A level with energy $\epsilon$ gives a contribution $\sim \delta/(\epsilon^2+\delta^2)$ to the LDOS $\rho$ at zero energy. It is easy to see that for small $\delta$ the zero-energy LDOS will be governed by the level closest to zero. Thus, we get
\be
\label{e8}
\langle \rho^q \rangle \propto \int d\epsilon [\delta/(\epsilon^2+\delta^2)]^q P(\epsilon),
\ee
where $P(\epsilon)$ is the distribution of the lowest energy level. For RM ensembles one has $P(\epsilon) \propto |\epsilon|^{m_l}$, where $m_l$ is the multiplicity of the long roots for the symmetric space of Hamiltonians. For all WD ensembles $m_l =0$, for class C $m_l=2$ and for class CI $m_l=1$. Equation (\ref{e8}) yields $\langle \rho^q\rangle \propto \delta^{m_l+1-q}$ for $q > (m_l+1)/2$ and $\langle \rho^q\rangle \propto \delta^q$ for $q < (m_l+1)/2$. This fixes the value of $q_*$ in  the relation (\ref{e7}):
\be
\label{e11}
q_* = m_l + 1 = \left\{
\begin{array}{ll}
1, & \qquad \text{WD classes}, \\
2, & \qquad \text{class CI}, \\
3, & \qquad \text{class C}.
\end{array}
\right.
\ee

We turn now to the derivation of Eq.~(\ref{e7}). The $\sigma$ mo\-dels are defined on symmetric superspaces $G/K$, where $G$ is a Lie supergroup and $K$ is a compact subgroup fixed by a Cartan involution \cite{SUSY, zirnbauer96, evers08}. We focus first on the unitary WD class (A); in this case the $\sigma$ model field $Q(r)= g(r)\sigma_3 g(r)^{-1}$ is a  $4\times 4$ supermatrix satisfying $Q^2=1$. Here $g(r) \in G$ and $\sigma_3$ is the third Pauli matrix in the retarded-advanced (RA) space. The moments of the LDOS at a point $r_0$ are given by \cite{mirlin94}
\be
\label{e12}
\langle \rho^q \rangle = \int \!\! DQ \Big[\frac{1}{2}\big(Q_{11} \!-\! Q_{22} \!+\! Q_{12} \!-\! Q_{21}\big)_{\rm bb}\Big]^q e^{-{\cal F}(Q)},
\ee
where $Q\equiv Q(r_0)$, with indices 1,2 referring to the RA decomposition and b,f to the boson-fermion one. The factor $e^{-{\cal F}(Q)}$ results from integrating out $Q(r)$ at the points $r\ne r_0$ and generically breaks the symmetry from $G$ to $K$ as a result of coupling the system to metallic reservoir(s). The only important property of the function ${\cal F}(Q)$ is its invariance with respect to the group $K$, i.e.\ ${\cal F}(kQk^{-1}) = {\cal F}(Q)$ for any $k\in K$. This follows from the corresponding invariance  of the action of the $\sigma$ model, including the boundary terms $\propto {\rm Str} \, \sigma_3 Q(r)$ appearing at points $r$ coupled to leads. In Ref.~\cite{mirlin94} the LDOS distribution function corresponding to (\ref{e12}) was evaluated by using the ``standard'' (introduced by Efetov) parametrization of the $Q$ field \cite{SUSY}, which led to Eq.~(\ref{e4}). In order to uncover the group-theoretic basis of the symmetry, we will use an alternative parametrization. It is based on the Iwasawa decomposition for symmetric superspaces \cite{mmz94, alldridge10} generalizing the corresponding construction for classical noncompact symmetric spaces \cite{helgason78}. By this decomposition, every element $g\in G$ is represented as $g = nak$ with $n \in N$, $a\in A$, and $k\in K$, where $A$ is a maximal Abelian subgroup for $G/K$ and $N$ is a nilpotent group. The decomposition is unique, once the set of positive roots is fixed. (The corresponding root vectors form the basis of the Lie algebra of $N$.)

It is convenient to switch to ${\mathcal Q} = Q\sigma_3$ and perform a unitary rotation ${\mathcal Q} \to \tilde{\mathcal Q} \equiv U{\mathcal Q}U^{-1}$ by the matrix $U=(1+i\sigma_1+i\sigma_2+i\sigma_3)/2$ in the RA space, which cyclically permutes Pauli matrices: $U\sigma_jU^{-1} = \sigma_{j-1}$. The combination of $Q_{ij}$ entering Eq.~(\ref{e12}) then becomes
\be
\label{e13}
(1/2)(Q_{11}-Q_{22}+Q_{12}-Q_{21})_{\rm bb} = \tilde{\mathcal Q}_{22,{\rm bb}}.
\ee
The Iwasawa decomposition of $g$ leads to ${\mathcal Q} = na^2\sigma_3n^{-1}\sigma_3$, where we used $k \sigma_3k^{-1} = \sigma_3$ and $a\sigma_3a^{-1} = a^2 \sigma_3$. Upon the rotation ${\mathcal Q} \to \tilde{\mathcal Q}$, this takes
the form $\tilde{\mathcal Q} = \tilde{n}\tilde{a}^2\sigma_2\tilde{n}^{-1}\sigma_2$, or explicitly
\be
\label{e14}
\tilde{\mathcal Q} =
\begin{pmatrix}
1 & * & * & * \\
0 & 1 & * & * \\
0 & 0 & 1 & 0 \\
0 & 0 & * & 1
\end{pmatrix}
\!\!
\begin{pmatrix}
e^{2x} & 0 & 0 & 0 \\
0 & e^{2iy} & 0 & 0 \\
0 & 0 & e^{-2x} & 0 \\
0 & 0 & 0 & \! e^{-2iy} \!
\end{pmatrix}
\!\!
\begin{pmatrix}
1 & 0 & 0 & 0 \\
* & 1 & 0 & 0 \\
* & * & 1 & * \\
* & * & 0 & 1
\end{pmatrix},
\ee
where $*$ denote some nonzero matrix elements of nilpotent matrices. The variables $x$ and $y$ (which correspond to $\lambda_1= \cosh \theta_1$ and $\lambda_2 = \sin\theta_2$ in the standard parametrization \cite{SUSY}) parametrize the Abelian group $A$. This group is noncompact in the $x$ direction and compact in the $y$ direction. It follows from (\ref{e14}) that the matrix element (\ref{e13}) is equal to  $\tilde{\mathcal Q}_{22,{\rm bb}} =
e^{-2x}$. The integral (\ref{e12}) for the LDOS moments thus becomes
\be
\label{e15}
\langle \rho^q \rangle = \int_{NA} \!\! Dn Da \, e^{-2\rho(\ln a)} e^{-2qx} e^{-{\cal F}(na^2\sigma_3 n^{-1}\sigma_3)},
\ee
where $Dn$ and $Da =dx dy$ are the invariant (Haar) measures on $N$ and $A$, respectively. The factor $e^{-2\rho(\ln a)}$ is the super-Jacobian, with $\rho(\ln a)$ being the half sum of positive roots; for the present case
\be
\label{e16}
\rho = -x + iy .
\ee

Next, we perform the $n$ integration involving only the last factor in the integrand of (\ref{e15}). For this purpose, we use the Harish-Chandra integral theorem stating that for a $K$-invariant function [$f(g)= f(kgk^{-1})$ for any $k\in K$]
\be
\label{e17}
\int_N dn f(na) = e^{\rho (\ln a)} I_f(a) ,\qquad I_f(a^w) = I_f(a).
\ee
The central point is that the function $I_f(a)$ is invariant with respect to the action $w:\ a\to a^w$ by any element $w \in W$ of the Weyl group $W$ of $G/K$. The classical version of the theorem (\ref{e17}) can be found, e.g., in \cite{helgason84}; the supersymmetric generalization (that we actually need) has been developed very recently \cite{alldridge10}. The Weyl group acts on the Lie algebra of $A$; its elements are reflections with respect to hyperplanes orthogonal to roots. The element that will be important for us here is the reflection $x \to -x$. Substituting Eqs.~(\ref{e16}) and  (\ref{e17}) into (\ref{e15}), we obtain
\be
\label{e18}
\langle \rho^q \rangle = \int dx dy \,  e^{(1-2q)x -iy } \, I_{\cal F}(x,y).
\ee
Finally, by using the symmetry of $I_{\cal F}(x,y)$ with respect to $x \to -x$, we obtain Eq.~(\ref{e4}).

Thus, the symmetry relation (\ref{e4}) previously derived for the WD classes is a consequence of the Weyl group invariance. We can now extend this result to two new classes, namely, C and CI. By inspecting the above derivation, one can see that the value of $q_*=1$ was determined by the coefficient in front of $x$ in the half sum of positive roots $\rho$, Eq.~(\ref{e16}). The analogous formulas for the classes C and CI read $\rho = -3x + 2iy_1 + iy_2$ and $\rho = -2x + 2iy$, reproducing the values of $q_*$ obtained from the RM argument, Eq.~(\ref{e11}).

What about the remaining five symmetry classes? The above derivation based on the Weyl group invariance is not directly applicable to them because of a more complicated structure of the $\sigma$ model target space. Specifically, that space contains an additional U(1) factor in the case of the chiral classes AIII, BDI, and CII, and a $\text{O}(1)=\mathbb{Z}_2$ factor for the classes D and DIII. More work is needed to explore the peculiar physics of Anderson localization in these symmetry classes.

The obtained symmetry relations are confirmed by a large body of analytical and numerical results for various disordered systems. For the WD classes, supporting evidence based on $2+\epsilon$ expansion and simulations of a power-law random banded matrix model was presented in Ref.~\cite{mirlin06}. Since then, a wealth of numerical
results on the 2D Anderson transition in the symplectic class AII \cite{symplectic}, the integer quantum Hall transition (unitary class A) \cite{qhe}, as well as the 3D transition in the orthogonal class AI \cite{vasquez08} have corroborated the relation (\ref{e3}).

A thoroughly investigated representative of symmetry  class C is the 2D spin quantum Hall (SQH) transition. For this system, it was proven analytically that, in the bulk, $x_\rho = 1/4$, $\Delta_2 = -1/4$, $\Delta_3 = -3/4$ \cite{gruzberg99, mirlin03}. In combination with the trivial values $\Delta_0=\Delta_1=0$ this yields $x_0 = x_3 = 0$, and $x_1 =x_2 = 1/4$, in agreement with the symmetry relation (\ref{e6}) with $q_* = 3$. Furthermore, these exponents have also been found for the SQH surface multifractality, with the results $x_0=x_3=0$ and $x_1=x_2=1/3$ \cite{subramaniam08}, again respecting the symmetry. Finally, the MF spectrum at the 2D SQH transition was studied numerically, both in the bulk and at the surface \cite{mirlin03, subramaniam06}. When expressed in terms of $x_q$, the data perfectly agree with the symmetry relation (\ref{e6}).

There also exists a model in class CI that has been studied in detail. This is the model of 2D Dirac fermions coupled to a random SU(2) gauge potential which is described by a Wess-Zumino-Witten theory \cite{su2gauge} and represents physics at the surface of a disordered 3D topological superconductor \cite{CITopSuperconductor}. Critical exponents for this model are known exactly: $x_\rho =1/4$ and $\tau_q = 2(q-1)(1-q/8)$  \cite{mudry96}, so that the LDOS MF spectrum reads $x_q = q(2-q)/4$, which clearly satisfies Eq. (\ref{e6}) with $q_* = 2$ [in agreement with (\ref{e11})].

Further support for our results is provided by the LDOS distribution of a quasi-1D system in contact with a metallic reservoir. Far away from the contact the distribution of $\ln \rho$ is known to be Gaussian, with the ratio ${\rm var}(\ln\rho)/\langle -\ln\rho\rangle$ equal to 2 for the WD classes, 2/3 for class C, and 1 for class CI \cite{beenakker97, brouwer00}. By calculating the moments $\langle \rho^q \rangle$, we recover Eqs.~(\ref{e7}, \ref{e11}) for these classes.

For the remaining five classes of Anderson localization, as well as for critical MF systems of other origin, our general arguments based on conformal symmetry predict a weaker (valid at criticality only) relation (\ref{e6}) to hold, with $q_*$ having the same degree of universality as critical exponents normally have (i.e.\ they are controlled by a particular fixed point rather than solely by the symmetry class). Several exactly solvable problems confirm this. In particular, the model of 2D Dirac fermions in an Abelian random vector potential (residing in chiral class AIII) has \cite{LFSG} a parabolic MF spectrum with the symmetry point $q_*/2 = 1/(4g_A)$ depending on the disorder strength $g_A$. The same applies to the non-Abelian SU($N$) (with $N\ne 2$) generalization of this model \cite{mudry96} (belonging to class AIII as well) where the symmetry point is $N$ dependent: $q_*/2 = N/(2N-2)$. Another example is the MF spectrum of the Ising disorder variable at the 2D Nishimori critical point, for which $q_*=1$ \cite{MerzChalker2002}. Furthermore, the MF spectra of the harmonic measure of critical curves \cite{duplantier00} can be obtained by introducing (conformal) primary operators characterized by charges $\alpha$ (analogous to our $q$) \cite{rushkin07}; their conformal weights $h_\alpha$ (analogs of our $x_q$) form a parabolic spectrum with a symmetry point $\alpha_*$ depending on the central charge of the model.

To summarize, conformal invariance arguments reveal the general character of the symmetry relation (\ref{e6}) for critical disordered systems with MF scaling of moments of observables (represented by local field operators ${\cal O}_q$). For the Anderson localization problem in the WD, C, and CI classes a stronger symmetry relation (\ref{e7}) holds, which is based on the Weyl group symmetry of the $\sigma$ model target space. This stronger relation is not restricted to criticality, and is characterized by a symmetry point $q_*/2$ depending on the symmetry class only. Future work should clarify the impact of the Weyl group symmetry on scaling dimensions of other composite operators, and explore the role of the U(1) and O(1) degrees of freedom in the chiral, D, and DIII classes.

The work was supported by DFG CFN (A. D. M), DFG SFB/TR 12 (M. R. Z), NSF Grants No. DMR-0448820 and No. DMR-0213745 (I. A. G.), and No. DMR-0706140 (A. W. W. L.). A. D. M. and A. W. W. L. acknowledge hospitality of the University of Chicago where part of this work was done.

\end{document}